%1234567890123456789012345678901234567890123456789012345678901234567890
\input phyzzx
\overfullrule=0pt
%\nopubblock
%\draft
%%%%%%%%%%%%%%%%%%%%%%%%%%%%%%%%%%%%%%%%%%%%%%%%%%%%%%%%%%%%%%%%%%%%%%
\def\psibar{\overline\psi}
\def\Dslash{D\kern-0.15em\raise0.17ex\llap{/}\kern0.15em\relax}
\def\Dslashl{\mathop{\Dslash}\limits^\leftarrow}
\def\partialmul{\mathop{\partial_\mu}\limits^\leftarrow}
\def\Rl{\mathop{R}\limits^\leftarrow}
%%%%%%%%%%%%%%%%%%%%%%%%%%%%%%%%%%%%%%%%%%%%%%%%%%%%%%%%%%%%%%%%%%%%%%
%
\REF\SCH{%
J. Schwinger,
Phys.\ Rev.\ {\bf 82}, 664 (1951).}
\REF\FUJ{%
K. Fujikawa,
Phys.\ Rev.\ {\bf D29}, 285 (1984);
Nucl.\ Phys.\ {\bf B428}, 169 (1994).}
\REF\BAN{%
H. Banerjee, R. Banerjee and P. Mitra,
Z. Phys.\ {\bf C32}, 445 (1986).}
\REF\SUZ{%
H. Suzuki,
Phys.\ Rev.\ {\bf D55}, 2994 (1997).}
\REF\KAR{%
L. H. Karsten and J. Smit,
Nucl.\ Phys.\ {\bf B183}, 103 (1981).}
\REF\NIE{%
H. B. Nielsen and M. Ninomiya,
Nucl.\ Phys.\ {\bf B185}, 20 (1981).}
\REF\WES{%
J. Wess and B. Zumino,
Phys.\ Lett. {\bf 37B}, 95 (1971).}
\REF\OKU{%
See, however, K. Okuyama and H. Suzuki,
Phys.\ Lett.\ {\bf B382}, 117 (1996), and references therein.}
\REF\WIL{%
K. G. Wilson,
{\it in\/} New Phenomena in Subnuclear Physics, ed.\ by
A. Zichichi (Plenum Press, New York, 1977).}
\REF\REB{%
C. Rebbi,
Phys.\ Lett.\ {\bf B186}, 200 (1987).}
\REF\NAR{%
R. Narayanan and H. Neuberger,
Nucl.\ Phys.\ {\bf B412}, 574 (1994);
{\it ibid.}\ {\bf B443}, 305 (1995);
{\it ibid.}\ {\bf B477}, 521 (1996);
Phys.\ Lett.\ {\bf B358}, 303 (1995);
{\it ibid.}\ {\bf B380}, 291 (1996).\nextline
R. Narayanan and P. Vranas,
Washinton University report, UW-PT-97-04, hep-lat/9702005.}
\REF\AOK{%
S. Aoki and R. B. Levien,
Phys.\ Rev.\ {\bf D51}, 3790 (1995).\nextline 
S. Randjbar-Daemi and J. Strathdee,
Phys.\ Lett.\ {\bf B348}, 543 (1995);
Phys.\ Rev.\ {\bf D51}, 6617 (1995);
Nucl.\ Phys.\ {\bf B443}, 386 (1995);
{\it ibid.}\ {\bf B461}, 305 (1996).\nextline
T. Kawano and Y. Kikukawa,
Kyoto University report, KUNS-1317, hep-lat/9501032.\nextline
M. F. L. Golterman and Y. Shamir,
Phys.\ Lett.\ {\bf B353}, 84 (1995);
{\it ibid.}\ {\bf B359}, 422 (1995) (E).\nextline
C. D. Fosco,
Int.\ J. Mod.\ Phys.\ {\bf A11}, 3987 (1996).\nextline
Y. Kikukawa and S. Miyazaki,
Prog.\ Theor.\ Phys.\ {\bf 96}, 1189 (1996).\nextline
Y. Kikukawa,
Kyoto University report, KUNS-1445, hep-lat/9705024.\nextline
N. Maru and J. Nishimura,
Nagoya University report, DPNU-97-25, hep-th/9705152.\nextline
A. Yamada, ICTP report, hep-lat/9705040.}
\REF\ZEN{%
S. V. Zenkin,
Phys.\ Lett.\ {\bf B395}, 283 (1997).}
\REF\BOD{%
G. T. Bodwin and E. V. Kov\'acs,
Nucl.\ Phys.\ (Proc.\ Suppl.) {\bf B30}, 617 (1993).\nextline
P. Hern\'andaz and R. Sundrum,
Nucl.\ Phys.\ {\bf B455}, 287 (1995);
{\it ibid.}\ {\bf B472}, 334 (1996).\nextline
G. T. Bodwin,
Phys.\ Rev.\ {\bf D54}, 6497 (1996).}
\REF\SEI{%
E. Seiler and I. Stamatescu,
Phys.\ Rev.\ {\bf D25}, 2177 (1982).}
\REF\PIE{%
S. D. Pietra and V. D. Pietra and L. Alvarez-Gaume,
Commun.\ Math.\ Phys.\ {\bf 109}, 691 (1987).\nextline
M. Gockeler and G. Schierholz,
Nucl.\ Phys.\ (Proc.\ Suppl.) {\bf B30}, 609 (1993),
and references therein.}
\REF\KAW{%
H. Kawai, R. Nakayama and K. Seo,
Nucl.\ Phys.\ {\bf B189}, 40 (1981).}
%%%%%%%%%%%%%%%%%%%%%%%%%%%%%%%%%%%%%%%%%%%%%%%%%%%%%%%%%%%%%%%%%%%%%%
\pubnum={IU-MSTP/21; hep-th/9706163}
\date={June 1997}
\titlepage
\title{Manifestly Gauge Covariant Treatment of Lattice Chiral
Fermions II}
\author{%
Kiyoshi Okuyama\foot{e-mail: okuyama@mito.ipc.ibaraki.ac.jp} and
Hiroshi Suzuki\foot{e-mail: hsuzuki@mito.ipc.ibaraki.ac.jp}}
\address{Department of Physics, Ibaraki University, Mito 310, Japan}
\abstract
%\doublespace
We propose a new formulation of chiral fermions on a lattice, on the
basis of a lattice extension of the covariant regularization scheme
in continuum field theory. The species doublers do not emerge. The
real part of the effective action is just one half of that of
Dirac-Wilson fermion and is always gauge invariant even with a finite
lattice spacing. The gauge invariance of the imaginary part, on the
other hand, sets a severe constraint which is a lattice analogue of
the gauge anomaly free condition. For real gauge representations, the
imaginary part identically vanishes and the gauge invariance becomes
exact.\nextline
\nextline
PACS number(s): 11.15.Ha, 11.30.Rd
\endpage
%%%%%%%%%%%%%%%%%%%%%%%%%%%%%%%%%%%%%%%%%%%%%%%%%%%%%%%%%%%%%%%%%%%%%%
%\doublespace
Inspired by the covariant regularization scheme~[\SCH,\FUJ,\BAN] in
the continuum field theory, one of us recently proposed a manifestly
gauge covariant treatment of chiral fermions on a lattice~[\SUZ].
However, the proposal was heavily relying on the notion in
perturbation theory and its validity was demonstrated only in the
continuum limit. Many important issues, such as the integrability
(see below), were also not clarified there. In this article, we
remedy these points and try to set up a truly non-perturbative
framework with the same strategy.

The basic idea of~[\SUZ] is the following: At present, it seems
impossible to construct a lattice action of chiral fermions which
{\it explicitly\/} distinguishes gauge anomaly free representations
{}from anomalous ones. This implies that we cannot expect a sensible
manifestly gauge invariant lattice formulation because it will not
reproduce in the continuum limit the gauge anomaly for the anomalous
cases. If one nevertheless forces the manifest gauge invariance, the
species doublers~[\KAR,\NIE], which cancel the gauge anomaly, will
emerge; thus we have to break the gauge symmetry at a certain stage.
With these observations, a formulation which preserves the gauge
symmetry as much as possible in {\it both\/} the anomalous and
non-anomalous cases seems desirable.

The covariant regularization~[\FUJ,\BAN] is such a regularization
scheme in the continuum theory. The scheme does not spoil {\it all\/}
the gauge invariance even in anomalous cases; instead it sacrifices
Bose symmetry among gauge vertices in a fermion one-loop diagram. In
this scheme, one starts with a regularized gauge current operator
(the covariant gauge current),
$$
\eqalign{
   \VEV{J^{\mu b}(x)}&=\VEV{\psibar(x)T^b\gamma^\mu P_R\psi(x)}
\cr
   &=-\lim_{y\to x}\tr T^b\gamma^\mu P_RG(x,y)
\cr
   &\equiv-\lim_{y\to x}\tr T^b\gamma^\mu P_Rf(\Dslash^2/\Lambda^2)
   {-1\over i\Dslash}\delta(x-y),
\cr
}
\eqn\one
$$
where $P_R\equiv(1+\gamma_5)/2$ is the chirality projection operator
and $\Dslash\equiv\gamma^\mu(\partial_\mu+iA_\mu^bT^b)$ is the
covariant derivative; note that {\it Dirac\/} propagator is used.
In~\one, $\Lambda$ is the cutoff parameter and the
regulating factor $f(t)$ satisfies $f(0)=1$ and
$f(\infty)=f'(\infty)=f''(\infty)=\cdots=0$. The definition
immediately follows the gauge {\it covariance\/} of the current
operator, namely, under the gauge transformation on the background
gauge field $A_\mu(x)\to%
-iV(x)\partial_\mu V^\dagger(x)+V(x)A_\mu(x)V^\dagger(x)$, the gauge
current transforms gauge covariantly
$$
   \VEV{J^{\mu b}(x)}
   \to-\lim_{y\to x}\tr [V^\dagger(x)T^bV(x)]
   \gamma^\mu P_Rf(\Dslash^2/\Lambda^2){-1\over i\Dslash}\delta(x-y).
\eqn\two
$$
In other words, the gauge invariance at external gauge vertices of a
fermion one loop diagram {\it except\/} that of $J^{\mu b}(x)$ is
preserved in the scheme. Because of this Bose asymmetric treatment of
gauge vertices, the gauge invariance can be ``maximally'' preserved
even in anomalous cases. As a consequence, the gauge anomaly has the
covariant form.

Once the gauge current operator is defined in this way, the effective
action~$\Gamma[A]$ might be obtained from the relation
$$
   \VEV{J^{\mu b}(x)}=-{\delta\Gamma[A]\over\delta A_\mu^b(x)}.
\eqn\three
$$
However such a functional $\Gamma[A]$ exists only if the covariant
gauge anomaly vanishes. The simplest way to see this is to note the
covariant anomaly does not satisfy Wess-Zumino consistency
condition~[\WES], which is a consequence of the integrability~\three.
The integrability or the Bose symmetry, however, is restored when we
can further impose the gauge invariance on
$J^{\mu b}(x)$-vertex, i.e., anomaly free cases. In fact, for anomaly
free cases, one can write down a formula of~$\Gamma[A]$~[\BAN],
$$
   \Gamma[A]=-\int_0^1dg\int d^4x\,A_\mu^b(x)\VEV{J^{\mu b}(x)}_g,
\eqn\four
$$
where the gauge current in the right hand side is the covariant
current~\one\ and the subscript~$g$ means it is evaluated by a
covariant derivative with a coupling constant~$g$,
$\Dslash_g\equiv\gamma^\mu(\partial_\mu+igA_\mu^bT^b)$. When the
gauge anomaly is absent, one can prove~[\BAN] that the integrable
current~\three\ coincides with the covariant one~\one\ (in the
infinite cutoff limit $\Lambda\to\infty$). In this scheme, therefore,
anomalous cases are distinguished by the non-integrability without
explicitly spoiling all the gauge invariance. 

The covariant current~\one\ is not in general integrable, i.e., not a
functional derivative of something. This means that in particular it
cannot be written as a functional derivative of the functional
integral of a certain action~[\OKU]. However one may directly work
with the fermion propagator and the gauge current operator as
in~\one. This is also true in the lattice theory; the crucial point
of our approach is to ``forget'' about the action~[\SUZ].

Let us now translate the above strategy of covariant regularization
into the lattice language as much as possible. Of course, there is a
wide freedom to do so, partially corresponding to the freedom of
regulating factor~$f(t)$. However the detail of the extension should
not be important and we first require followings: 1)~The expression
reduces to the continuum analogue in the naive (or classical)
continuum limit. 2)~The lattice propagator has no doubler's pole.
3)~The lattice fermion propagator transforms gauge covariantly,
namely, under the gauge transformation on the link variable
$U_\mu(x)\to V(x)U_\mu(x)V^\dagger(x+a^\mu)$,
the propagator transforms as $G(x,y)\to V(x)G(x,y)V^\dagger(y)$.

For definiteness and for simplicity, we will use Wilson
propagator~[\WIL] in this article:
$$
   G(x,y)\equiv{-1\over i\Dslash(x)+R(x)}\delta(x,y)
   =\delta(x,y){1\over i\Dslashl(y)+\Rl(y)},
\eqn\five
$$
where the delta function on the lattice is defined by
$\delta(x,y)\equiv\delta_{x,y}/a^4$; $\Dslash(x)$ is the lattice
covariant derivative and $R(x)$ is Wilson term:
$$
\eqalign{
   &\Dslash(x)\equiv\sum_\mu\gamma^\mu{1\over2a}
   \bigl[U_\mu(x)e^{a\partial_\mu}
                   -e^{-a\partial_\mu}U_\mu^\dagger(x)\bigr],
\cr
   &R(x)\equiv{r\over2a}\sum_\mu
   \bigl[U_\mu(x)e^{a\partial_\mu}
                   +e^{-a\partial_\mu}U_\mu^\dagger(x)-2\bigr],
\cr
}
\eqn\six
$$
and
$$
\eqalign{
   &\Dslashl(x)\equiv-\sum_\mu\gamma^\mu{1\over2a}
   \bigl[U_\mu(x)e^{-a\partialmul}
                   -e^{a\partialmul}U_\mu^\dagger(x)\bigr],
\cr
   &\Rl(x)\equiv-{r\over2a}\sum_\mu
   \bigl[U_\mu(x)e^{-a\partialmul}
                   +e^{a\partialmul}U_\mu^\dagger(x)-2\bigr].
\cr
}
\eqn\sixadd
$$
In the above expressions, $a$~is the lattice spacing and
$\exp(\pm a\partial_\mu)$ is the translation operator to
$\mu$-direction by a unit lattice spacing. The equality of two
expressions in~\five\ follows from two equivalent forms of Wilson
action,
$$
   S[\psi,\psibar,U]
   =a^4\sum_x\psibar(x)\bigl[i\Dslash(x)+R(x)\bigr]\psi(x)
   =-a^4\sum_x\psibar(x)\bigl[i\Dslashl(x)+\Rl(x)\bigr]\psi(x).
\eqn\seven
$$
In contrast to the continuum Dirac propagator in~\one, the Wilson
term mixes the right handed and left handed chiralities.\foot{%
One may even avoid this chiral symmetry breaking by making use of
more ingenious propagator in~[\REB]. See~[\SUZ].}
However we do not think this is so problematical because anyway the
physical particle picture emerges only in the continuum limit and, in
the continuum limit, we expect this chirality mixing due to the
Wilson term vanishes. Note that the Wilson propagator nevertheless
has the required gauge covariance property.

As the lattice analogue of the covariant gauge current, therefore
we shall study following object:
$$
\eqalign{
   \Delta[U,\delta U]&\equiv
   -a^4\sum_x\tr\bigl[i\delta\Dslash(x)P_R+{1\over2}\delta R(x)\bigr]
   G(x,y)\Bigr|_{y=x}
\cr
   &=a^4\sum_x\tr G(y,x)
   \bigl[i\delta\Dslashl(x)P_R+{1\over2}\delta \Rl(x)\bigr]
   \Bigr|_{y=x},
\cr
}
\eqn\eight
$$
where $\delta U$ represents an infinitesimal variation of the link
variable and its conjugate is defined by
$\delta U_\mu^\dagger(x)%
=-U_\mu^\dagger(x)\delta U_\mu(x)U_\mu^\dagger(x)$. The second
expression follows from the definitions \six\ and~\sixadd\ and the
fact that we can freely shift the ``integration variable''~$x$. As
the analogue of~\three, we identify it with the variation of the
effective action:
$$
   \Delta[U,\delta U]=\delta\Gamma[U].
\eqn\nine
$$
The defining relations, \eight\ and~\nine, are suggested by the
{\it naive\/} relation:
$\exp\Gamma[U]=\int{\cal D}\psi{\cal D}\psibar%
\exp[a^4\sum_x\psibar(x)i\Dslash(x)P_R\psi(x)]$. [The variation of
Wilson term $\delta R(x)$ in~\eight\ will be necessary for the
integrability.] The integrability~\nine\ is of course not a trivial
statement and will be investigated below.

We first note the manifest gauge covariance of $\Delta[U,\delta U]$:
$$
   \Delta\bigl[V(x)U_\mu(x)V^\dagger(x+a^\mu),\delta U_\mu(x)\bigr]
   =\Delta\bigl[U_\mu(x),V^\dagger(x)\delta U_\mu(x)V(x+a^\mu)\bigr].
\eqn\ten
$$
That is, $\Delta[U,\delta U]$ behaves gauge covariantly under the
gauge transformation on the background~$U$. This is an analogous
relation to~\two.

Next, we separate the ``would-be variation''~$\Delta[U,\delta U]$
into the real and imaginary parts. We note relations hold for an
arbitrary matrix~$m(x)$,
$$
   \Dslash(x)^*m(x)=-\bigl[m(x)^T\Dslashl(x)\bigr]^T,\quad
   R(x)^*m(x)=-\bigl[m(x)^T\Rl(x)\bigr]^T,
\eqn\eleven
$$
where $T^{b*}=T^{bT}$, $\gamma^{\mu*}=-\gamma^{\mu T}$ and
$\gamma_5^*=\gamma_5^T$ have been used. Using these relations, we
find,
$$
   G(x,y)^*=\gamma_5^TG(y,x)^T\gamma_5^T.
\eqn\twelve
$$
{}From \eleven\ and~\twelve, the complex conjugate
of~$\Delta[U,\delta U]$ is given by
$$
\eqalign{
   \Delta[U,\delta U]^*&
   =a^4\sum_x\tr G(y,x)
   \bigl[i\delta\Dslashl(x)P_L+{1\over2}\delta\Rl(x)\bigr]
   \Bigr|_{y=x}
\cr
   &=-a^4\sum_x\tr\bigl[i\delta\Dslash(x)P_L
                               +{1\over2}\delta R(x)\bigr]
   G(x,y)\Bigr|_{y=x}.
\cr
}
\eqn\thirteen
$$
Then a comparison with~\eight\ shows the real and imaginary parts are
respectively given by
$$
   {\rm Re}\,\Delta[U,\delta U]
   =-{1\over2}a^4\sum_x
   \tr\bigl[i\delta\Dslash(x)+\delta R(x)\bigr]G(x,y)\Bigr|_{y=x},
\eqn\fourteen
$$
and
$$
\eqalign{
   i\,{\rm Im}\,\Delta[U,\delta U]
   &=-{1\over2}a^4\sum_x\tr
   i\delta\Dslash(x)\gamma_5G(x,y)\Bigr|_{y=x}
\cr
   &={1\over2}a^4\sum_x
   \tr G(y,x)i\delta\Dslashl(x)\gamma_5\Bigr|_{y=x}.
\cr
}
\eqn\fourteenadd
$$

Now, for the {\it real\/} part of~$\Delta[U,\delta U]$~\fourteen, we
immediately see the integrability and the gauge {\it invariance}. By
the gauge invariance, we mean that the ``would-be variation'' of the
effective action, $\Delta[U,\delta U]$, vanishes along the direction
of the gauge degrees of freedom. That is,
$$
   {\rm Re}\,\Delta[U,\delta_\lambda U]=0,\quad{\rm for}\quad
   \delta_\lambda U_\mu(x)\equiv
   -i\lambda(x)U_\mu(x)+iU_\mu(x)\lambda(x+a^\mu),
\eqn\fifteen
$$
where $\lambda(x)=\lambda^b(x)T^b$. One can easily verify this
relation by using above definitions. This gauge invariance property
of the real part is almost trivial in our construction, because
${\rm Re}\,\Delta[U,\delta_\lambda U]$ is simply one half of that of
Dirac-Wilson fermion:
$$ 
   {\rm Re}\,\Delta[U,\delta U]=\delta\Gamma_1[U],\quad
   \Gamma_1[U]\equiv{1\over2}\ln\det\bigl[i\Dslash(x)+R(x)\bigr].
\eqn\sixteen
$$
Note that the last expression is well-defined and not a formal
one with the lattice regularization. Therefore, for the real part,
we arrived at a quite simple picture: The real part
of~$\Delta[U,\delta U]$ can always be regarded as a variation of the
effective action~$\Gamma_1[U]$, which is just one half of the effective
action of Dirac-Wilson fermion. In other words, the chiral
determinant obtained by
``integrating''~${\rm Re}\,\Delta[U,\delta U]$ gives rise to the
square root of Dirac-Wilson determinant. Although the gauge
invariance of the real part of the effective action is almost trivial
in this way, this seems very interesting because the gauge invariance
of the real part is one of main achievements of recent
researches~[\NAR,\AOK,\ZEN,\BOD]. In our approach, the origin of this
nice behavior of the real part may be traced to the basic idea of
covariant regularization, i.e., maximal gauge invariance. We note that
our treatment of the real part turned out to be almost identical to
that of~[\BOD].

The gauge invariance of the imaginary part, on the other hand, is
difficult. A short calculation shows,
$$
   i\,{\rm Im}\,\Delta[U,\delta_\lambda U]=a^4\sum_x\lambda^b(x)
   {\cal A}^b(x),
\eqn\seventeen
$$
where ${\cal A}^b(x)$ is given by
$$
   {\cal A}^b(x)\equiv
   -{1\over2}\tr\bigl[G(y,x)\Dslashl(x)\gamma_5T^b
                      -T^b\gamma_5\Dslash(x)G(x,y)\bigr]\Bigr|_{y=x}.
\eqn\eighteen
$$
In fact this is a lattice analogue of the gauge anomaly: By
considering the {\it axial\/} rotation
$\psi(x)\to\exp[i\theta^b(x)T^b\gamma_5]\psi(x)$ and
$\psibar(x)\to\psibar(x)\exp[i\theta^b(x)T^b\gamma_5]$ in the
Wilson action~\seven, we can compute~${\cal A}^b(x)$ in
the continuum limit~[\KAR,\SEI] and find the covariant gauge anomaly
$$
   \lim_{a\to0}{\cal A}^b(x)=
   {i\over32\pi^2}
   \varepsilon^{\mu\nu\rho\sigma}\tr T^bF_{\mu\nu}F_{\rho\sigma}.
\eqn\nineteen
$$
Therefore, if the gauge representation is anomaly free, the imaginary
part of $\Delta[U,\delta U]$ vanishes along the gauge variation
{\it in the continuum limit} and the effective action becomes gauge
invariant; this is the expected property. However this is not
sufficient for the gauge invariance with a {\it finite\/} lattice
spacing. It is clear that ${\cal A}^b(x)=0$ with a finite lattice
spacing is a much stronger condition than the anomaly free condition
in the continuum theory. We can furthermore show that the
integrability of the imaginary part also requires ${\cal A}^b(x)=0$
(see Appendix), thus the integrability does not hold
unless~${\cal A}^b(x)=0$.

Therefore we again face the usual difficulty of lattice chiral gauge
theory that the gauge mode decouples only in the continuum limit,
even in anomaly free cases. Although the natural lattice extension of
the covariant regularization provides a simple picture for a treatment
of the real part of the effective action, it does not solve the main
difficulty of anomaly free {\it complex\/} representations in the
lattice chiral gauge theory. For the general discussion on the
imaginary part of the effective action of lattice chiral fermion,
see~[\PIE]. Eq.~\nineteen\ suggests that the
difficulty of our approach might be avoided only by invoking the
double-limit procedure in~[\BOD].

However at least for {\it real\/} gauge representations, we can show
the above problems of the gauge invariance and the integrability do
not occur at all. This is because the imaginary part
of~$\Delta[U,\delta U]$~\fourteenadd\ identically vanishes for real
representations. The demonstration is straightforward: For a real
representation~$T^b$, there exists a unitary matrix~$u$ which
maps~$T^b$ into the conjugate representation,
$uT^bu^\dagger=-T^{b*}=-T^{bT}$. We then insert
$u^\dagger C^{-1}Cu=1$ into the first line of~\fourteenadd. ($C$ is
the charge conjugation matrix,
$C\gamma^\mu C^{-1}=-\gamma^{\mu T}$ and thus
$C\gamma_5C^{-1}=\gamma_5^T$.) Then by noting
$$
   Cu\delta\Dslash(x)u^\dagger C^{-1}m(x)
   =-\bigl[m(x)^T\delta\Dslashl(x)\bigr]^T,\quad
   CuG(x,y)u^\dagger C^{-1}=G(y,x)^T,
\eqn\twenty
$$
we find
$$
   i\,{\rm Im}\,\Delta[U,\delta U]
   =-{1\over2}a^4\sum_x
   \tr G(y,x)i\delta\Dslashl(x)\gamma_5\Bigr|_{y=x}.
\eqn\twentyadd
$$
A comparison with~\fourteenadd\ shows the imaginary part
of~$\Delta[U,\delta U]$ identically vanishes; $\Delta[U,\delta U]$
is purely real.

Therefore, the treatment of real representations is simple: The
variation of the effective action is given by~\fourteen, which is
nothing but the half of that of Dirac-Wilson fermion. We note
that, although this seems almost trivial, the square root of Dirac
determinant in general cannot be expressed as a functional integral
of a local action. In particular, it seems impossible to construct a
gauge invariant Wilson action for an odd number of chiral fermions in
a pseudo-real representation. The expression of the variation of
effective action~\fourteen\ furthermore seems congenial to Metropolis
algorithm, in which the {\it difference\/} of the effective action
between two gauge field configurations is the basic building block.
Thus we propose the use of~\fourteen. We have also established the
reality of the variation that is required in Metropolis algorithm. Of
course, since eq.~\fourteen\ represents only an infinitesimal change
of the effective action, presumably one has to divide a finite
variation associated with the update of a link variable into
sufficiently many pieces.

Relating to the actual numerical application, we have to investigate
also the necessity of the fine tuning. Although usually Wilson fermion
requires the fine tuning to restore the chiral symmetry~[\KAR], we do
not see the necessity in our formula~\fourteen: The configuration of
the link variable is kept fixed when computing the
variation~$\Delta[U,\delta U]$ and the original Wilson
propagator~\five\ as it stands is used. Therefore, for us, it seems
that the ``back-reaction'' of the gauge field dynamics does not
modify the above properties.

The overlap formulation~[\NAR,\AOK] also possesses nice properties
such that the real part of the effective action is gauge invariant
and that there is no need of the fine tuning. However, the overlap
has the remarkable property~[\NAR] that a relation of non-trivial
topological gauge field configurations and the fermionic zero mode is
explicit. In our approach, an investigation on such a
``global property'' has to be postponed as a future work.

Finally we comment on the relation to the continuum theory. By
parameterizing the link variable as $U_\mu(x)=\exp[iaA_\mu^b(x)T^b]$,
the gauge current is defined by
$$
\eqalign{
   &\VEV{J^{\mu b}(x)}
   \equiv-{\Delta[U,\delta U]\over a^4\delta A_\mu^b(x)}
\cr
   &=-\tr\int_0^1d\beta\,e^{\beta iaA_\mu(x)}
   T^be^{-\beta iaA_\mu(x)}
\cr
   &\quad\times
   {1\over2}\Bigl[
   \bigl(\gamma^\mu P_R-{ir\over2}\bigr)U_\mu(x)G(x+a^\mu,x)
   +\bigl(\gamma^\mu P_R+{ir\over2}\bigr)
   G(x,x+a^\mu)U_\mu^\dagger(x)\Bigr].
\cr
}
\eqn\twentyone
$$
The fermion one loop vertex functions are defined accordingly,
$$
\eqalign{
   \VEV{J^{\mu b}(x)}
   &\equiv\sum_{n=1}^\infty{1\over n!}\prod_{j=1}^n
   \Bigl[
   a^4\sum_{x_j,\mu_j,b_j}A_{\mu_j}^{b_j}(x_j)
   \int_{-\pi/a}^{\pi/a}
   {d^4p_j\over(2\pi)^4}\,e^{ip_j(x-x_j)}e^{-ia{p_j}_{\mu_j}/2}
   \Bigr]
\cr
\noalign{\vskip2pt}
   &\qquad\qquad\qquad\times
 \Gamma^{\mu\mu_1\cdots\mu_n\,bb_1\cdots b_n}(p_1,p_2,\cdots,p_n).
\cr
}
\eqn\twentytwo
$$
When a new lattice formulation is proposed, it is important to examine
the continuum limit in the perturbative treatment. However, in our
formulation, the real part of the gauge current~\twentyone\ is just
one half of that of the conventional Wilson fermion. Therefore, for
the real part, Ward identities associated with the gauge
symmetry~[\KAR], which are linear relations among vertex functions,
trivially hold. Also all the perturbative calculations for the vertex
functions of Wilson fermion can be used by simply dividing by two.
For example, we may use the result of~[\KAW] for the vacuum
polarization tensor (because of~$\gamma_5$, the imaginary part does
not contribute to this function) to yield,
$$
   \lim_{a\to0}\Gamma^{\mu\nu\,bc}(p)
   =-{1\over24\pi^2}\tr T^bT^c(p^\mu p^\nu-g^{\mu\nu}p^2)
   \Bigl[\log{4\pi\over-a^2p^2}
         -\gamma+{5\over3}-12\pi^2L(r)\Bigr],
\eqn\twentythree
$$
where the function $L(\lambda)$ is given by Eq.~(3.25) of~[\KAW].

For the imaginary part of the gauge current~\twentyone, our
construction~\fourteenadd\ is quite faithful to the idea of covariant
regularization. For example, using the gauge covariance~\ten, we can
derive Ward identities associated with the gauge invariance at
external vertices~[\SUZ]:
$$
\eqalign{
   &p_\nu\lim_{a\to0}\Gamma^{\mu\nu\,bc}(p)=0,
\cr
   &p_\nu\lim_{a\to0}\Gamma^{\mu\nu\rho\,bcd}(p,q)
   +if^{bce}\lim_{a\to0}\Gamma^{\mu\rho\,ed}(q)
   -if^{cde}\lim_{a\to0}\Gamma^{\mu\rho\,be}(p+q)=0,
\cr
}
\eqn\twentyfour
$$
and so on. Eq.~\nineteen\ on the other hand shows we have the
covariant gauge anomaly, which completely vanishes for anomaly free
cases without any gauge non-invariant counter terms. Therefore,
assuming the Lorentz covariance is restored, we can expect the
continuum limit of our formulation reproduces all the results of the
covariant regularization in the continuum theory.

We thank K.~Haga for collaboration in the early stage. We are
grateful to Prof.~K.~Fujikawa for discussions and to
Prof.~H.~Banerjee and Prof.~P.~Mitra for answering our question
on~[\BAN]. We are also grateful to Prof.~S.~V.~Zenkin for helpful
information. The work of H.S. was supported in part by the Ministry
of Education Grant-in-Aid for Scientific Research, Nos.~09740187,
09226203 and 08640348.

\appendix

In this Appendix, we present a relation between the integrability
of the imaginary part of~$\Delta[U,\delta U]$ and the ``anomaly free
condition'' ${\cal A}^b(x)=0$. First we define a quantity:
$$
   K^\mu(x)_{ij}\equiv{i\,{\rm Im}\,\Delta[U,\delta U]\over
                       a^4\delta U_\mu(x)_{ji}}.
\eqn\aone
$$
We perform the infinitesimal gauge transformation $\delta_\lambda U$
in~\fifteen\ in the both sides of this equation. In the left hand
side, the gauge transformation may be generated by a differential
operator,
$$
   G^b(y)\equiv
   \sum_\nu\Bigl\{
   -i\bigl[T^bU_\nu(y)\bigr]_{lk}{\delta\over\delta U_\nu(y)_{lk}}
   +i\bigl[U_\nu(y-a^\nu)T^b\bigr]_{lk}
   {\delta\over\delta U_\nu(y-a^\nu)_{lk}}
   \Bigr\}
\eqn\atwo
$$
It is easy to see that $\sum_x\lambda^b(x)G^b(x)$ generates the
infinitesimal gauge transformation. Then we can cast the gauge
variation of the left hand side into the following form:
$$
\eqalign{
   &G^b(y)K^\mu(x)_{ij}
\cr
   &={\delta\over\delta U_\mu(x)_{ji}}
   {\cal A}^b(y)
\cr
   &\quad+i\bigl[K^\mu(x)T^b\bigr]_{ij}\delta_{x,y}
    -i\bigl[T^bK^\mu(x)\bigr]_{ij}\delta_{x+a^\mu,y}
\cr
   &\quad+\sum_\nu\Bigl\{
   -i\bigl[T^bU_\nu(y)\bigr]_{lk}{\cal R}^{\mu\nu}_{ij,kl}(x,y)
   +i\bigl[U_\nu(y-a^\nu)T^b\bigr]_{lk}
    {\cal R}^{\mu\nu}_{ij,kl}(x,y-a^\nu)
   \Bigr\}.
\cr
}
\eqn\athree
$$
In deriving this identity, we first interchanged the places of $ij$
and~$kl$. This produced a ``functional rotation'' of $K$,
$$
   {\cal R}^{\mu\nu}_{ij,kl}(x,y)\equiv
   {\delta K^\mu(x)_{ij}\over\delta U_\nu(y)_{lk}}
   -{\delta K^\nu(y)_{kl}\over\delta U_\mu(x)_{ji}}.
\eqn\afour
$$
We then changed the order of the derivative and $U$. This produced
the commutator term in the second line of~\athree.

Now, the right hand side of~\aone\ transforms gauge covariantly under
the infinitesimal gauge transformation. This can be written as
$$
   G^b(y){i\,{\rm Im}\,\Delta[U,\delta U]\over
                       a^4\delta U_\mu(x)_{ji}}
   =i\bigl[K^\mu(x)T^b\bigr]_{ij}\delta_{x,y}
    -i\bigl[T^bK^\mu(x)\bigr]_{ij}\delta_{x+a^\mu,y}.
\eqn\afive
$$
Therefore, from \athree\ and~\afive, we find
$$
\eqalign{
   &{\delta\over\delta U_\mu(x)_{ji}}{\cal A}^b(y)
\cr
   &=\sum_\nu\Bigl\{
   i\bigl[T^bU_\nu(y)\bigr]_{lk}{\cal R}^{\mu\nu}_{ij,kl}(x,y)
   -i\bigl[U_\nu(y-a^\nu)T^b\bigr]_{lk}
   {\cal R}^{\mu\nu}_{ij,kl}(x,y-a^\nu)
   \Bigr\}.
\cr
}
\eqn\asix
$$
The right hand side of this equation can be regarded as the covariant
divergence of the functional rotation~${\cal R}$. We can interpret
this identity from two different view points. First, if the lattice
gauge anomaly ${\cal A}^b(x)$ vanishes, then the covariant divergence
of the functional rotation~${\cal R}$ vanishes. A similar relation
to~\asix\ exists in the continuum theory and, when the gauge anomaly
is absent, it can be used to show the functional rotation of the
covariant gauge current vanishes. This fact was used to show the
integrability of the covariant current in anomaly free cases~[\BAN].
In our present lattice case, unfortunately, we could not prove that
the corresponding statement that the covariant conservation, i.e.,
$\hbox{\asix}=0$, implies the vanishing of~${\cal R}$. {\it If\/} the
functional rotation~\afour\ itself is zero, then Poincar\'e's lemma
may be used to show the (local) integrability of the imaginary part,
$$
   i\,{\rm Im}\,\Delta[U,\delta U]=\delta\Gamma_2[U].
\eqn\aseven
$$

On the contrary, if we assume the integrability~\aseven, we have
${\cal R}=0$ and~\asix\ shows ${\cal A}^b(x)$ is independent of $U$.
However we can directly compute ${\cal A}^b(x)$ for $U=1$
with a finite lattice spacing and find ${\cal A}^b(x)=0$ for $U=1$.
Consequently, the integrability requires the lattice anomaly free
condition, ${\cal A}^b(x)=0$.

\refout
\bye